\documentstyle[aps,pre,multicol,epsfig]{revtex}

\begin{document} \bibliographystyle{unsrt} 
\input{epsf}  
\title{Memory and Chaos Effects in Spin Glasses}
\author{K. Jonason$^2$, E. Vincent$^1$, J. Hammann$^1$, J.P. Bouchaud$^1$, P. Nordblad$^2$}
\address{$^1$Service de Physique de l'Etat Condens\'e, CEA Saclay,\\ 91191 Gif sur Yvette Cedex, France}
\address{$^2$Department of Material Science, Uppsala University, P.O. Box
534, \\ 751 21 Uppsala, Sweden }
\maketitle

\begin{abstract}

New low frequency $ac$ susceptibility measurements on two different spin
glasses show that cooling/heating the sample at a constant rate yields
an essentially reversible (but rate dependent) $\chi(T)$ curve; a downward
relaxation of $\chi$ occurs during a temporary stop at constant
temperature ({\it ageing}).  Two main features of our results are: (i)
when cooling is resumed after such a stop, $\chi$ goes back to the
reversible curve ({\it chaos}) (ii) upon re-heating, $\chi$ perfectly
traces the previous ageing history ({\it memory}). We discuss
implications of our results for a {\it real space} (as opposed to {\it
phase space}) picture of spin glasses.

\end{abstract}

\centerline{PACS numbers: 75.50.Lk 75.10.Nr}
\vskip 0.2cm
{\it \centerline{ to appear in Phys. Rev. Lett.}}

\bigskip
\begin{multicols}{2}

\narrowtext

The dynamic properties of the spin glass phase have been extensively
studied by both experimentalists and theorists for almost two decades
\cite{Young,saclayrev1}.  The observed properties reflect the
out-of-equilibrium state of the system: the response to a field
variation is logarithmically slow, and, in addition, depends on the
time spent at low temperature (``ageing''). Ageing is fully
reinitialized by heating the sample above the glass temperature
$T_g$. It corresponds to the slow evolution of the system towards
equilibrium, starting at the time of the quench below $T_g$. Many
aspects of ageing are similar to the``physical ageing'' phenomena that
have been characterized in the mechanical properties of glassy
polymers \cite{struik}. In the last few years, some interesting
progress in the theoretical understanding of ageing in disordered
systems has been achieved \cite{mfnonequ}.

From the studies of the critical behaviour at $T_g$ \cite{critic}, it
appears that the approach of $T_g$ is associated to the divergence of
a spin-spin correlation length, as is the case in the phase transition
of classical ordered systems.  In the spin glass phase, the system is
out of equilibrium: as in simple ferromagnets, it is tempting to
associate ageing with the progressive growth of a typical domain size
towards an equilibrium infinite value. However, this simple picture
cannot account for all the experimental observations.  In particular,
the effect of small temperature cycles (within the spin-glass phase)
is rather remarkable \cite{cycleuppsala,cyclesaclay}:

$\bullet$ on the one hand, ageing at a higher temperature barely contributes
to ageing at a lower temperature.  Said differently (as will be
discussed again below), the thermal history at sufficiently higher
temperatures is irrelevant. This is at variance with a simple scenario
of thermal activation over barriers, where the time spent at higher
temperature would obviously help the system to find its equilibrium
state. Everything happens as if there were strong changes of the
free-energy landscape with temperature. This point is suggestive of
the ``chaotic'' aspect of the spin glass phase that has been predicted
from mean field theory \cite{mpv} and from scaling arguments in
\cite{BrayMoore,FHlett}.

$\bullet$ on the other hand, interesting memory effects concomitantly appear:
the state reached by the system at a given temperature can be
retrieved after a negative temperature cycle.

In the present letter, we describe some new experiments which reveal
in a rather striking way these memory and chaos effects, and we point
out their implications for the construction of a real space picture of
spin glasses.  The results are obtained using a new experimental
protocol, which has first been proposed and applied to the metallic
Cu:Mn spin glass by one of us \cite{Nordblad}. We now develop this
approach in a series of measurements on the $CdCr_{1.7}In_{0.3}S_4$
insulating spin glass \cite{mtrlref}. The universality of the
out-of-equilibrium dynamics in spin glasses is evidenced by the
similarity between the results on two very different realisations of
spin glass systems.

In this procedure, we record the ac-susceptibility of the sample as a
function of temperature.  The ac field has a low frequency of
$\omega/2\pi=0.04 Hz$, to allow the relaxation of the susceptibility
due to ageing to be clearly visible on the scale of several hours.
The peak amplitude of the ac field is $0.3Oe$, which is low enough not
to affect the properties of the system. We cool the system from above
$T_g=16.7K$ down to $5K$ at a constant cooling rate of $0.1K/min$, and
then heat it back continuously at the same rate. This yields two
slightly different curves; the one obtained upon heating (a bit below
the other one) is chosen as the {\it reference curve}, and shown as a
solid line in Fig. 1.  We then repeat the experiment, but now stop
during cooling at an intermediate temperature $T_{1}=12 K$ during a
certain waiting time $t_{w1}=7 h$.  During $t_{w1}$, due to ageing,
both $\chi'$ and $\chi''$ relax downwards, by about the same amount;
for $\chi''$, however, the relative amount is much larger, which makes
the effect more visible, and in the following we mainly concentrate on
the out-of-phase component. After the ageing stage at $T_1$, the
cooling procedure resumes, and one observes that $\chi''$ and $\chi'$
merge back with the reference curve only a few Kelvin below $T_1$.
Thus, ageing at $T_1$=12 K has not influenced the result at lower
temperatures ({\it ``chaos'' effect}).

\vskip 1.2cm
\begin{figure}

\centerline{\hbox{\epsfig{figure=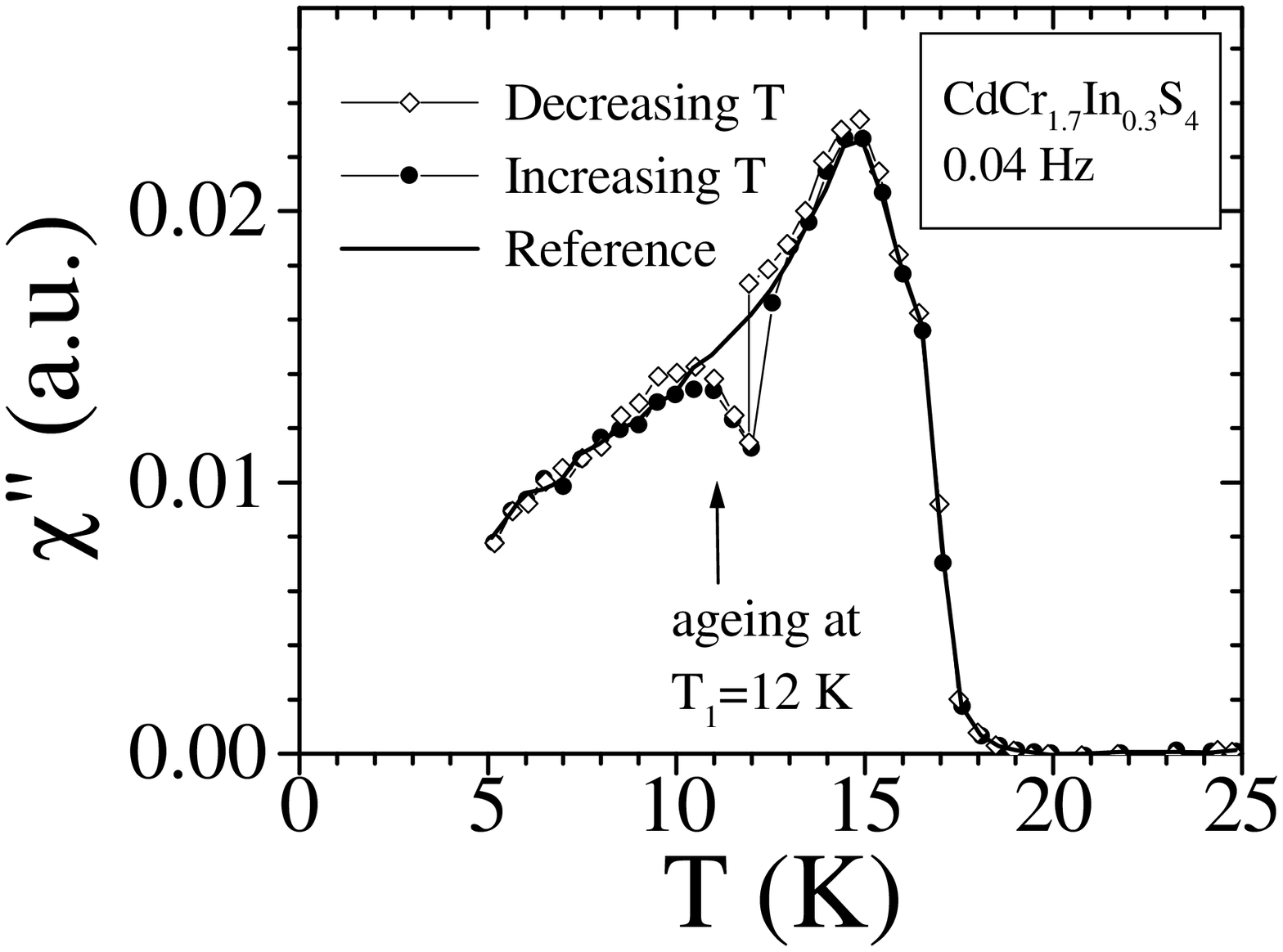,width=8.5cm}}}
\caption{
\hbox{Out-of-phase susceptibility $\chi''$ of the}
$CdCr_{1.7}In_{0.3}S_4$ spin glass.  The solid line is measured upon
heating the sample at a constant rate of $0.1K/min$ (reference
curve). Open diamonds: the measurement is done during cooling at this
same rate, except that the cooling procedure has been stopped at $12K$
during $7h$ to allow for ageing. Cooling then resumes down to $5K$:
$\chi''$ is not influenced and goes back to the reference curve
(``chaos''). Full circles: after this cooling procedure, the data is
taken while re-heating at the previous constant rate, exhibiting
memory of the ageing stage at $12K$.
}
\label{fig1}
\end{figure}

The surprise is that when the sample is re-heated at a constant
heating rate (i.e. no further stops on the way up), we find that the
trace of the previous stop (the dip in $\chi''$) is exactly recovered
(see Fig 1). The memory of what happened at $T_{1}=12 K$ has not been
erased by the further cooling stage, {\it even though $\chi''$ at
lower temperatures lies on the reference curve}. The system can
actually retrieve information from several stops if they are
sufficiently separated in temperature. In Fig.2, we show a ``double
memory experiment'', in which two ageing evolutions, one at $T_1=12K$
and the other at $T_2=9K$, are retrieved \cite{remark}. In the inset
of Fig.2, the result of a similar experiment on a Cu:Mn sample is
shown \cite{Nordblad}.

As discussed above, the cooling rate dependence of the dynamics in
spin glasses is largely governed by the ``chaos'' effect.  For
example, it has been shown that there is no difference in the ageing
behaviour if the spin glass has been directly quenched from above $T_g$
or if it has been subjected to a very long waiting pause immediately
below $T_g$ \cite{cyclesaclay}.  However, the influence of the cooling
rate was
not quantitatively characterized in systematic measurements, and
this point is of a particular interest for the comparison between spin
glasses and other glassy systems.  We have therefore performed the
following experiment. We cool the sample progressively and
continuously (in fact, by steps of $0.5K$) from above $T_g$ to
$12K=0.72T_g$, using three very different cooling rates. The result is
shown in Fig.3.  The initial values of $\chi'$ and $\chi"$ are indeed
different: slower cooling yields a smaller initial value of the
susceptibility, a value that is closer to ``equilibrium''.  A small
horizontal shift of the curves along the time scale allows the
superposition of the three of them; the curves obtained after a slower cooling are somewhat ``older''. However, all curves are clearly converging
towards the same asymptotic value. This behaviour contrasts with that
observed in systems where thermally activated domain growth is important, for
example the dielectric relaxation of the dipole glass
$K_{1-x}Li_{x}TaO_3$ \cite{levelut}. There, it is found that different
cooling rates lead the system to very different apparent asymptotic values of
the dielectric constant.  

\vskip 1.5cm
\begin{figure}
\centerline{
\hbox{\epsfig{figure=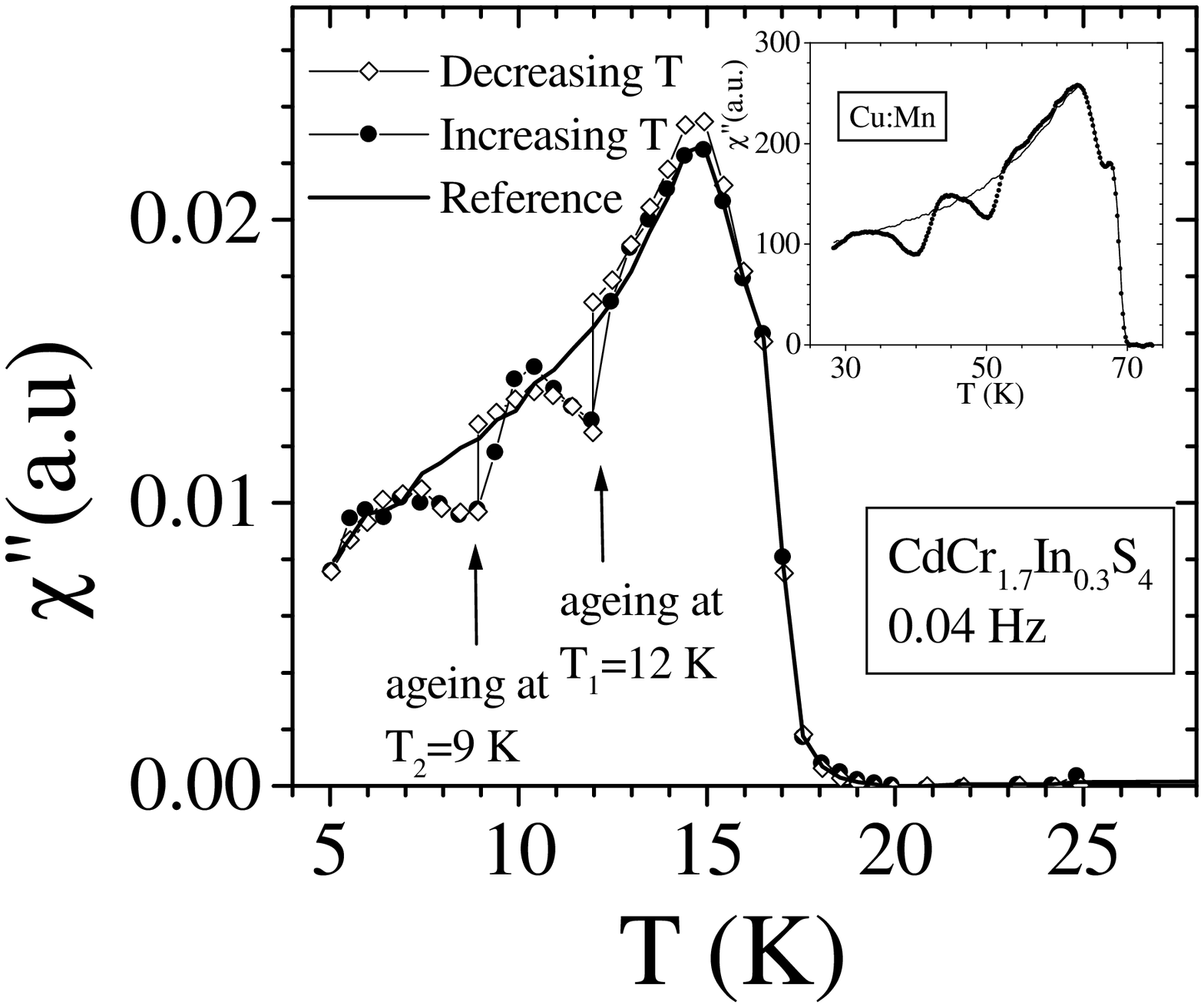,width=8.5cm}}
}

\caption{
Same as in Fig.1 ($CdCr_{1.7}In_{0.3}S_4$ insulating sample), 
but with two stops during cooling, which allow the
spin glass to age $7h$ at $12K$ and then $40h$ at $9K$. Both ageing
memories are retrieved independently when heating back (full circles).
The inset shows a similar ``double memory'' experiment performed on the
Cu:Mn metallic spin glass [11].
} 
\label{fig2}
\end{figure}

One can furthermore show that the cooling rate effect seen in Fig. 3
is entirely due to the last temperature interval, and not at all to
the time spent at higher temperatures. We again use different cooling
rates from above $T_g=16.7K$ to $14K$, but then we rapidly cool from
$14K$ to $12K$, where the relaxation is measured. In this procedure,
despite very different average cooling rates, the last two Kelvin are
always crossed at the same speed. The result, in Fig. 3 (inset), is
unambiguous: the obtained relaxation is the same in all cases, for
$\chi'$ as well as for $\chi''$. Thus, in a spin glass, the only
influence of the cooling rate on the ageing state is due to the very
last temperature interval before reaching the measurement temperature,
while due to chaos effects the time spent at higher temperatures does
not contribute. Again, this strongly contrasts with the case of
$K_{1-x}Li_{x}TaO_3$ alluded to above, where it is the time spent at
temperatures near the glass transition that mostly determines the
final state after the quench \cite{levelut}.

\vskip 0.7cm
\begin{figure}
\centerline{\hbox{\epsfig{figure=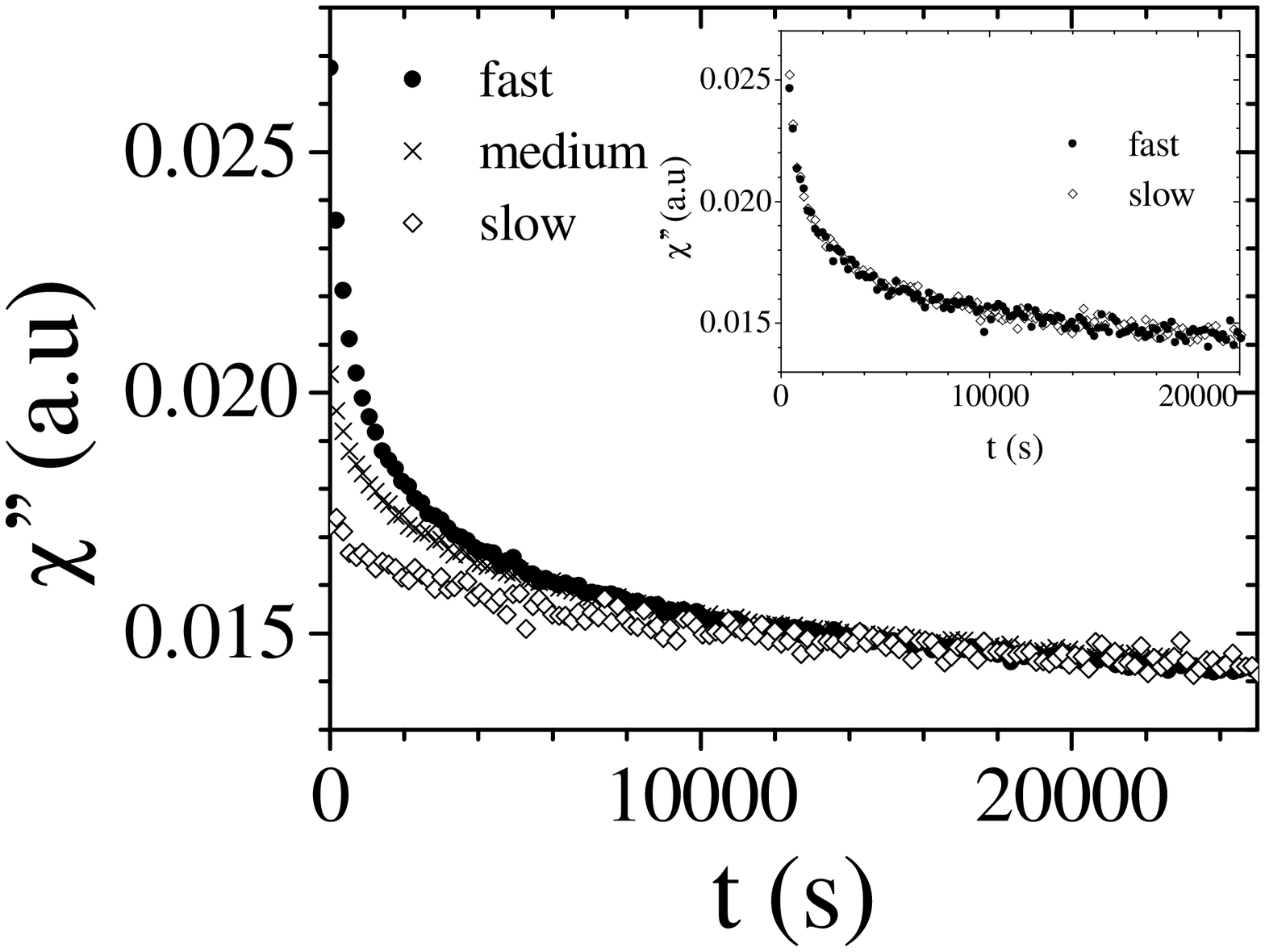,width=8.5cm}}}
\vskip 0.5cm\caption{
$\chi''$ relaxation at $12K$ as a
function of time: effect of the cooling rate on ageing. The 
$CdCr_{1.7}In_{0.3}S_4$ sample has been cooled from above $T_g=17K$ to
$12K$ at very different speeds: $2.6K/min$ (full circles),
$0.08K/min$ (crosses), $0.015K/min$ (open diamonds). 
In the inset, 
another procedure is used which shows that this cooling rate effect is
only due to the last temperature interval:
constant rate of $0.8K/min$ (full circles) or $0.08K/min$ (open
diamonds) from $17K$ to $14K$, but in both cases rapid quench 
from $14K$ to $12K$.
} 
\label{fig3}
\end{figure}

There has been quite a number of approaches, inspired by Parisi's
solution of mean-field models, in which ageing can be described in
terms of a random walk in the space of the metastable states
\cite{sibani,jpbdean,orbach}.  The memory and chaos effects have been
described in terms of a {\it hierarchical organization of the
metastable states as a function of temperature} \cite{cyclesaclay}, a
picture in which the growth of the free energy barriers when the
temperature decreases could be characterized quantitatively
\cite{sacorbach}.  Although these phase space pictures are very
helpful (and have been used very early in the context of spin-glasses
and glasses), it is obvious that they need to be linked with real
space pictures, where the experimental signal can be attributed to
certain clusters of spins which flip collectively. Macroscopically,
ageing means that the system becomes ``stiffer'' with time, in the
sense that the response to a field variation becomes slower and slower
with increasing age. Slower response means larger free-energy
barriers, and correspondingly a larger number of spins to be
simultaneously flipped: one is thus naturally led to think in terms of
growing ``domains'' (or ``droplets'') of correlated spins. The
simplest picture based on this idea has been proposed by Fisher and
Huse \cite{FH} and Koper and Hilhorst \cite{KH} in slightly different
terms.  It is based on the postulate that, at any given temperature
below the spin-glass transition, there is only one phase (and its spin
reversed counterpart) to be considered, much as in a standard
ferromagnet. The difference is of course that all spins do not point
in one direction, but arrange in a random (but fixed) way imposed by
the interplay between the disordered nature of the interactions and
the temperature. One can however by convention call one of the phases
`up' and the other one `down'; again as in a ferromagnet, the typical
domain size is expected to grow with time, albeit logarithmically
slowly since domain walls are pinned by the disorder.

In principle, this picture should lead to very strong cooling rate
effects, which are, as stated above, not those that are observed
experimentally in spin glasses (see Fig. 3). However, if one assumes
(as suggested by mean field and scaling arguments) {\it chaos with
temperature}, in the sense that the phase growing at temperature $T_1$
is not at all the ``correct'' equilibrium phase for another
temperature $T_2$, the cooling rate dependence can indeed be small
since the time spent at a higher temperature does not bring the system
any closer to equilibrium. This must be contrasted with random field
like systems, where the equilibrium state is the same in the whole low
temperature phase, and where cooling rate effects are strong.

The chaotic dependence of the phase with temperature allows one to
argue why ageing is restarted when the temperature is lowered: the
configuration reached after ageing at $T_1$ is, from the point of view
of the `$T_2$-phase', completely random. Correspondingly, new
$T_2$-domains have to grow. The problem, however, comes from the
observed memory effect: it shows that the $T_2$-domains must indeed
grow somewhere, but {\it without destroying the preexisting
$T_1$-domains}. The only possibility is that the $T_2$-domains {\it do
not nucleate everywhere}, but only around certain favorable nucleation
centers, {\it coexisting} with the previous backbone of $T_1$-domains.
This is however in contradiction with the idea that, at any
temperature, only one phase (and thus two types of domains) is enough
to describe the dynamics of the system completely, since we already
require the coexistence of {\it two} types of domains ($T_1$- and
$T_2$-domains). The same argument shows that, at the first temperature
$T_1$, the system must actually be in a mixture of all the different
phases encountered between $T_g$ and $T_1$. Intuitively, it seems
clear that if the system is so fragile to temperature changes, then by
the same token it is hard to imagine that other nearby `phases' will
not nucleate simultaneously with the nominal phase. In other words,
the question is whether, at long times, the `defects' are only domain
walls between two well identified phases (as in the coarsening/droplet
picture), or whether these defects are more complicated (branched)
objects.

We believe that the memory effect discussed above is incompatible with
a picture of the {\it out-of-equilibrium} dynamics based on one type
of domains only. Obviously, this does not exclude the possibility that
the {\it equilibrium} phase is unique; from an experimental point of
view, however, the question is not relevant, since the system is never
in equilibrium. A similar conclusion has been suggested by `second
noise spectrum' experiments \cite{Weissman}, and by recent out of
equilibrium numerical simulations \cite{Marinari}.

The construction of a consistent `hierarchical droplet' picture
appears completely open and beyond the scope of the present paper. A
possibility (discussed in\cite{jpbdean,mfnonequ}) is that correlations
at a given temperature establish themselves progressively, but only
over non compact (i.e. fractal) clusters of spins, which are large
enough to be frozen at that temperature, but leave the surrounding sea
of spins relatively free. As the temperature is lowered, smaller
clusters begin to freeze (and thus provide a new aging signal), while
larger clusters are completely blocked (thus leading to the memory
effect).  Let us note that the idea of ``droplets within droplets''
has previously been discussed in \cite{Villain,fractclust} without (to
our knowledge) getting to the stage of a more quantitative model.  The
present experiments, which suggest the coexistence of many different
time scales (and thus, presumably, many length scales) may force one
to take the idea of fractal droplets more seriously.

\vskip 0.5cm
\noindent
{\it Acknowledgements} Financial support from the Swedish Natural Science
Research Council (NFR) is acknowledged; one of us (KJ) wants to thank
the Swedish Institute (SI) for a fellowship. We are grateful to
F. Alberici, L.F. Cugliandolo, P. Doussineau, A. Levelut, M. M\'ezard,
M. Ocio, G. Parisi for enlightening discussions, and to L. Le Pape for
his technical support.

\end{multicols} 


\begin{thebibliography}{99}
\bibitem{remark}
Note that if $T_2$ is very close to $T_1$, ageing at $T_2$ slowly
wipes out the memory of the dip at $T_1$. On the other hand, both dips
are erased by heating above $T_1$.
\end{thebibliography}
\end{document}